# Pricing the Aunt Michaela Option with a Modified Black-Scholes Equation with a Maturity Condition of Gamma Type

Juan Ospina
Computational Quantum Econo-physics Group
Mathematical Physics Program
Manrique Institute of Technology (MIT)
Medellín, Colombia
jospina@gmail.com

**Abstract.**

Using Maple, we compute a new exact series solution of a modified Black-Scholes equation, recently proposed, for the case of the Aunt Michaela option with a maturity condition of gamma type. We show that the modified Black-Scholes equation with the Aunt Michaela option is exactly solvable in terms of associated Laguerre polynomials or equivalently, in terms of Whittaker M functions. Finally, we make some numerical experiments with the analytical solutions .

**Keywords:** Computational quantum econo-physics, modified Black-Scholes equation, pricing options, Aunt Michaela option, special functions, Whittaker functions, Laguerre polynomials, symbolic computation, computer algebra, Maple.

## 1. Introduction

The standard Black-Scholes equation [1,2,3]

$$\left(\frac{\partial}{\partial t} V(t,S)\right) + r S \left(\frac{\partial}{\partial S} V(t,S)\right) + \frac{1}{2}\sigma^2 S^2 \left(\frac{\partial^2}{\partial S^2} V(t,S)\right) - r V(t,S) = 0 \tag{0.1}$$

provides a tool to estimate the prices of different speculative financial options. According with the standard Black-Scholes equation (0.1) the price $V(t,S)$ of a speculative financial option is a function of the time $t$ and the current price $S$ of the underlying asset; and such price $V(t,S)$ depends on the volatility $\sigma$ and the risk-free interest rate $r$. From the mathematical point of view, according with the equation (0.1), the volatility $\sigma$ plays the role of a diffusivity; and the risk-free interest rate $r$ plays simultaneously the role of drift parameter and reaction rate.

In order to obtain an explicit solution of the standard Black-Scholes equation (0.1) it is necessary to specify the kind of speculative financial option that we are considering. There are two classical kinds of speculative financial options. The first one is called *European call option* and it is defined by [1]

$$V(T,S) = \begin{cases} S - K & K \leq S \\ 0 & S < K \end{cases} \tag{0.2}$$

where *K* is named the strike price of the underlying asset and *T* is the maturity time for the considered speculative financial option. Then, a first mathematical problem consists in to solve (0.1) with the condition (0.2).

The second classical kind of speculative financial option is called *European put option* which is defined by [1]

$$V(T,S) = \begin{cases} 0 & K \leq S \\ K - S & S < K \end{cases} \tag{0.3}$$

Then, a second mathematical problem consists in to solve (0.1) with the condition (0.3).

It is well known that the solution of (0.1) with (0.2) has the form [1]

$$V_{call}(t,S) = S\left(\frac{1}{2} + \frac{1}{2}\mathrm{erf}\left(\frac{1}{2}\frac{\sqrt{2}\left(\ln\left(\frac{S}{K}\right) + \left(r + \frac{\sigma^2}{2}\right)(T-t)\right)}{\sigma\sqrt{T-t}}\right)\right)$$
$$- K\,\mathrm{e}^{(-r(T-t))}\left(\frac{1}{2} + \frac{1}{2}\mathrm{erf}\left(\frac{1}{2}\frac{\sqrt{2}\left(\ln\left(\frac{S}{K}\right) + \left(r - \frac{\sigma^2}{2}\right)(T-t)\right)}{\sigma\sqrt{T-t}}\right)\right) \tag{0.4}$$

and the solution of (0.1) with (0.3) is given by [1]

$$V_{put}(t,S) = -S\left(\frac{1}{2} - \frac{1}{2}\mathrm{erf}\left(\frac{1}{2}\frac{\sqrt{2}\left(\ln\left(\frac{S}{K}\right) + \left(r + \frac{\sigma^2}{2}\right)(T-t)\right)}{\sigma\sqrt{T-t}}\right)\right)$$
$$+ K\,\mathrm{e}^{(-r(T-t))}\left(\frac{1}{2} - \frac{1}{2}\mathrm{erf}\left(\frac{1}{2}\frac{\sqrt{2}\left(\ln\left(\frac{S}{K}\right) + \left(r - \frac{\sigma^2}{2}\right)(T-t)\right)}{\sigma\sqrt{T-t}}\right)\right) \tag{0.5}$$

Using (0.4) and (0.5) it is possible to compute the theoretical prices of many speculative financial options when the relevant parameters are known.

In the speculative "real" life there are appreciable deviations of "real" prices of the speculative financial options respect to the theoretical prices prescribed by the solutions of the Black-Scholes equation. In order to close the gap between the "observed" prices of the speculative financial options and the theoretical prices, some authors have proposed two kinds of strategies: a) formulate generalized Black-Scholes equations which contain the standard Black-Scholes equation as a particular case [1]; b) formulate modified Black-Scholes equations which not necessary contain the standard Black-Scholes equation as a particular case [3].

In this work we consider the case of a modified Black-Scholes equation recently proposed [3,3A] and we will compute certain exact series solution in the case of the *Aunt Michaela option* with a maturity condition given in terms a gamma distribution. For the computation we will use computer algebra software, specifically Maple [4] and we will use some special functions of the mathematical physics such as the Whittaker functions [5] and the associated Laguerre polynomials [6].

## 2. Problem

We consider here the modified Black-Scholes equation proposed recently by Y. Zheng, namely

$$\left(\frac{\partial}{\partial t} V(t, S)\right) + r S \left(\frac{\partial}{\partial S} V(t, S)\right) + \frac{1}{2} \sigma(S, t)^2 \left(\frac{\partial^2}{\partial S^2} V(t, S)\right) - r V(t, S) = 0 \qquad (1)$$

When

$$\sigma(S, t) = \sigma S \qquad (2)$$

the equation (1) is reduced to the classical Black-Scholes equation

$$\left(\frac{\partial}{\partial t} V(t, S)\right) + r S \left(\frac{\partial}{\partial S} V(t, S)\right) + \frac{1}{2} \sigma^2 S^2 \left(\frac{\partial^2}{\partial S^2} V(t, S)\right) - r V(t, S) = 0 \qquad (3)$$

In this paper we study a family of particular cases of (1) with

$$\sigma(S, t) = \sigma S^{\left(\frac{k}{2}\right)} \qquad (4)$$

When (4) is replaced in (1) we obtain the sub-Black-Scholes equation with the form

$$\left(\frac{\partial}{\partial t} V(t, S)\right) + r S \left(\frac{\partial}{\partial S} V(t, S)\right) + \frac{1}{2} \sigma^2 S^k \left(\frac{\partial^2}{\partial S^2} V(t, S)\right) - r V(t, S) = 0 \qquad (5)$$

where (5) is reduced to (3) for $k = 2$.

Our problem here consists is to solve the equation (5) with the *Aunt Michaela option* given by

$$V(S, T) = \frac{A S^{(p+1)} e^{(-\alpha S)} \alpha^{(p+1)}}{\Gamma(p+1)} \qquad (6)$$

where $p$, $A$ and $\alpha$ are the parameters that are specifying the Aunt Michaela option.

## 3. Method

In this section a computational procedure will be given step by step in order to solve the problem (5)-(0.3). We will use Maple and we will apply some special functions of the mathematical physics such as the Whittaker functions and the associated Laguerre polynomials. Specifically we will use the orthonomality property of the associated Laguerre polynomials, which has the form [7]

$$\int_0^\infty e^{(-x)} x^k L(n, k, x) L(m, k, x) \, dx = \frac{(n+k)! \, \delta_{n, m}}{n!} \qquad (7A)$$

where $L(n,k,x)$ is the associated Laguerre polynomial with degree $n$ and order $k$ in the variable $x$. For the computations will be use the Maple notation

$$\text{LaguerreL}(n, k, x) = L(n, k, x) \tag{7B}$$

We look for a solution of the equation (5) with the form

$$C(x, t) = e^{(-\lambda t)} F(x) \tag{8}$$

Replacing (8) in (5) we obtain

$$\left(\frac{\partial}{\partial t}(e^{(-\lambda t)} F(S))\right) + r S \left(\frac{\partial}{\partial S}(e^{(-\lambda t)} F(S))\right) + \frac{1}{2} \sigma^2 S^k \left(\frac{\partial^2}{\partial S^2}(e^{(-\lambda t)} F(S))\right) - r e^{(-\lambda t)} F(S)$$
$$= 0$$

$$\tag{9}$$

which is reduced to

$$-\lambda F(S) + r S \left(\frac{d}{dS} F(S)\right) + \frac{1}{2} \sigma^2 S^k \left(\frac{d^2}{dS^2} F(S)\right) - r F(S) = 0 \tag{10}$$

According with Maple the solution of (10) is given by

$$F(S) = e^{\left(\frac{S^{(-k+2)} r}{(k-2) \sigma^2}\right)} S^{\left(\frac{k}{2} - 1/2\right)} \left(\_C1 \text{ WhittakerM}\left(\frac{-2\lambda - 3r + rk}{2 r (k-2)}, \frac{1}{2(k-2)}, \frac{2 S^{(-k+2)} r}{(k-2) \sigma^2}\right)\right.$$
$$\left. + \_C2 \text{ WhittakerW}\left(\frac{-2\lambda - 3r + rk}{2 r (k-2)}, \frac{1}{2(k-2)}, \frac{2 S^{(-k+2)} r}{(k-2) \sigma^2}\right)\right)$$

$$\tag{11}$$

Given that we need a bounded solution when $S = 0$; and given that the function WhittakerW tends to $\infty$ when $S = 0$, we demand that $\_C2 = 0$; and then (11) is reduced to

$$F(S) = \_C1\, e^{\left(\frac{S^{(-k+2)} r}{(k-2) \sigma^2}\right)} S^{\left(\frac{k}{2} - 1/2\right)} \text{WhittakerM}\left(\frac{-2\lambda - 3r + rk}{2 r (k-2)}, \frac{1}{2(k-2)}, \frac{2 S^{(-k+2)} r}{(k-2) \sigma^2}\right)$$

$$\tag{12}$$

The solution (12) can be rewritten as

$$F(S) = \_C1\, \text{LaguerreL}\left(\frac{-r - \lambda}{(k-2) r}, \frac{1}{k-2}, \frac{2 S^{(-k+2)} r}{(k-2) \sigma^2}\right) 2^{\left(\frac{k-1}{2(k-2)}\right)} (k-2)^{\left(-\frac{k-1}{2(k-2)}\right)}$$
$$r^{\left(\frac{k-1}{2(k-2)}\right)} \sigma^{\left(-\frac{k-1}{k-2}\right)} \Big/ \text{binomial}\left(-\frac{\lambda}{(k-2) r}, -\frac{r+\lambda}{(k-2) r}\right)$$

$$\tag{13}$$

In order to guaranty that the associated *LaguerreL* function in (13) will be a polynomial we demand that

$$\frac{-r-\lambda}{(k-2)r} = n \qquad (14)$$

where $n$ is a natural number. From (14) we deduce that

$$\lambda = -n(k-2)r - r \qquad (15)$$

Replacing (15) in (13) and redefining the integration constant, we get

$$F(S) = C\,\text{LaguerreL}\left(n, \frac{1}{k-2}, \frac{2\,S^{(-k+2)}\,r}{(k-2)\,\sigma^2}\right) \qquad (16)$$

Given that the modified Black-Scholes equation (5) is a linear equation, it is possible to apply the superposition principle; and then, using (16) and (8) with (15), the general solution of (5) takes the form

$$V(t,S) = \sum_{n=0}^{\infty} c_n \,\text{LaguerreL}\left(n, \frac{1}{k-2}, \frac{2\,S^{(-k+2)}\,r}{(k-2)\,\sigma^2}\right) e^{(r(nk-2n+1)t)} \qquad (17)$$

We solve (5) with the Aunt Michaela option given

$$V(S,T) = \frac{A\,S^{(p+1)}\,e^{(-\alpha S)}\,\alpha^{(p+1)}}{\Gamma(p+1)} \qquad (18)$$

and for hence, from (17) and (18) we have that

$$V(T,S) = \sum_{n=0}^{\infty} c_n \,\text{LaguerreL}\left(n, \frac{1}{k-2}, \frac{2\,S^{(-k+2)}\,r}{(k-2)\,\sigma^2}\right) e^{(r(nk-2n+1)T)} \qquad (19)$$

Making the change of variable $u = \dfrac{2\,S^{(-k+2)}\,r}{(k-2)\,\sigma^2}$, it is to say $S = \left(\dfrac{u\,\sigma^2\,(k-2)}{2\,r}\right)^{\left(\frac{1}{-k+2}\right)}$

the equation (19) takes the form

$$\frac{A\,S^{(p+1)}\,e^{(-\alpha S)}\,\alpha^{(p+1)}}{\Gamma(p+1)} = \sum_{n=0}^{\infty} c_n \,\text{LaguerreL}\left(n, \frac{1}{k-2}, u\right) e^{(r(nk-2n+1)T)} \qquad (19A)$$

Now, multiplying the both sides of (19A) by $\text{LaguerreL}\left(m, \dfrac{1}{k-2}, u\right) u^a\,e^{(-u)}$ we get

$$\frac{A\,S^{(p+1)}\,e^{(-\alpha S)}\,\alpha^{(p+1)}}{\Gamma(p+1)}\,\mathrm{LaguerreL}\!\left(m,\frac{1}{k-2},u\right) u^a\,e^{(-u)} =$$

$$\sum_{n=0}^{\infty} c_n\,\mathrm{LaguerreL}\!\left(n,\frac{1}{k-2},u\right)\mathrm{LaguerreL}\!\left(m,\frac{1}{k-2},u\right) u^a\,e^{(-u)}\,e^{(r(nk-2n+1)T)}$$

(20)

where $a$ is certain constant to be determined.

Integrating the both sides of (20) respect to $S$ from 0 to $\infty$, the equation (20) is transformed to

$$\int_0^\infty \frac{A\,S^{(p+1)}\,e^{(-\alpha S)}\,\alpha^{(p+1)}}{\Gamma(p+1)}\,\mathrm{LaguerreL}\!\left(m,\frac{1}{k-2},u\right) u^a\,e^{(-u)}\,dS =$$

$$\sum_{n=0}^{\infty} c_n \int_0^\infty \mathrm{LaguerreL}\!\left(n,\frac{1}{k-2},u\right)\mathrm{LaguerreL}\!\left(m,\frac{1}{k-2},u\right) u^a\,e^{(-u)}\,dS\,e^{(r(nk-2n+1)T)}$$

(21)

Making the variable change given by $u = \dfrac{2\,S^{(-k+2)}\,r}{(k-2)\,\sigma^2}$ , in the integral on the right hand side of (21) we obtain

$$\int_0^\infty \frac{A\,S^{(p+1)}\,e^{(-\alpha S)}\,\alpha^{(p+1)}}{\Gamma(p+1)}\,\mathrm{LaguerreL}\!\left(m,\frac{1}{k-2},u\right) u^a\,e^{(-u)}\,dS =$$

$$\sum_{n=0}^{\infty} c_n \int_0^\infty \mathrm{LaguerreL}\!\left(n,\frac{1}{k-2},u\right)\mathrm{LaguerreL}\!\left(m,\frac{1}{k-2},u\right) u^{\left(\frac{ak-2a+1-k}{k-2}\right)} e^{(-u)}\,2^{\left(\frac{1}{k-2}\right)} r^{\left(\frac{1}{k-2}\right)}$$

$$\sigma^{\left(-\frac{2}{k-2}\right)}(k-2)^{\left(-\frac{k-1}{k-2}\right)}du\,e^{(r(nk-2n+1)T)}$$

(22)

In order for reducing the integral in the right hand side of (22) to the orthonormalization relation of the associated Laguerre polynomials given by (7A) we demand that $\dfrac{ak-2a+1-k}{k-2} = \dfrac{1}{k-2}$ ; which implies that $a = \dfrac{k}{k-2}$. Replacing such value of $a$ in the right hand side of (22) we obtain

$$\int_0^\infty \frac{A\,S^{(p+1)}\,e^{(-\alpha S)}\,\alpha^{(p+1)}}{\Gamma(p+1)}\,\mathrm{LaguerreL}\!\left(m,\frac{1}{k-2},u\right) u^{\left(\frac{k}{k-2}\right)} e^{(-u)}\,dS = \sum_{n=0}^\infty c_n \int_0^\infty$$

$$\mathrm{LaguerreL}\!\left(n,\frac{1}{k-2},u\right) \mathrm{LaguerreL}\!\left(m,\frac{1}{k-2},u\right) u^{\left(\frac{1}{k-2}\right)} e^{(-u)}\, 2^{\left(\frac{1}{k-2}\right)} r^{\left(\frac{1}{k-2}\right)} \sigma^{\left(-\frac{2}{k-2}\right)}$$

$$(k-2)^{\left(-\frac{k-1}{k-2}\right)} du\, e^{(r(nk-2n+1)T)}$$

(22A)

Then, using (7A) in the integral of the right hand side of (22A) we have that

$$\int_0^\infty \mathrm{LaguerreL}\!\left(n,\frac{1}{k-2},u\right) \mathrm{LaguerreL}\!\left(m,\frac{1}{k-2},u\right) u^{\left(\frac{1}{k-2}\right)} e^{(-u)}\, 2^{\left(\frac{1}{k-2}\right)} r^{\left(\frac{1}{k-2}\right)} \sigma^{\left(-\frac{2}{k-2}\right)}$$

$$(k-2)^{\left(-\frac{k-1}{k-2}\right)} du = \frac{2^{\left(\frac{1}{k-2}\right)} r^{\left(\frac{1}{k-2}\right)} \sigma^{\left(-\frac{2}{k-2}\right)} (k-2)^{\left(-\frac{k-1}{k-2}\right)} \left(\frac{nk-2n+1}{k-2}\right)!\,\delta_{n,m}}{n!}$$

(23)

Replacing (23) in (22A) and using $u = \dfrac{2\,S^{(-k+2)}\,r}{(k-2)\sigma^2}$ ; it gives that

$$\int_0^\infty \frac{A\,S^{(p+1)}\,e^{(-\alpha S)}\,\alpha^{(p+1)}}{\Gamma(p+1)}\,\mathrm{LaguerreL}\!\left(m,\frac{1}{k-2},\frac{2\,S^{(-k+2)}\,r}{(k-2)\sigma^2}\right)\left(\frac{2\,S^{(-k+2)}\,r}{(k-2)\sigma^2}\right)^{\left(\frac{k}{k-2}\right)} e^{\left(-\frac{2\,S^{(-k+2)}\,r}{(k-2)\sigma^2}\right)} dS =$$

$$\sum_{n=0}^\infty \frac{c_n\, 2^{\left(\frac{1}{k-2}\right)} r^{\left(\frac{1}{k-2}\right)} \sigma^{\left(-\frac{2}{k-2}\right)} (k-2)^{\left(-\frac{k-1}{k-2}\right)} \left(\frac{nk-2n+1}{k-2}\right)!\,\delta_{n,m}\, e^{(r(nk-2n+1)T)}}{n!}$$

(24)

which is reduced to

$$\int_0^\infty \frac{A\,S^{(p+1)}\,e^{(-\alpha S)}\,\alpha^{(p+1)}}{\Gamma(p+1)}\,\mathrm{LaguerreL}\!\left(m,\frac{1}{k-2},\frac{2\,S^{(-k+2)}\,r}{(k-2)\sigma^2}\right)\left(\frac{2\,S^{(-k+2)}\,r}{(k-2)\sigma^2}\right)^{\left(\frac{k}{k-2}\right)} e^{\left(-\frac{2\,S^{(-k+2)}\,r}{(k-2)\sigma^2}\right)} dS =$$

$$\frac{c_m\, 2^{\left(\frac{1}{k-2}\right)} r^{\left(\frac{1}{k-2}\right)} \sigma^{\left(-\frac{2}{k-2}\right)} (k-2)^{\left(-\frac{k-1}{k-2}\right)} \left(\frac{mk-2m+1}{k-2}\right)!\, e^{(r(mk-2m+1)T)}}{m!}$$

(25)

From (25) we derive that

$$m! \frac{\int_0^\infty \frac{A S^{(p+1)} e^{(-\alpha S)} \alpha^{(p+1)}}{\Gamma(p+1)} \text{LaguerreL}\left(m, \frac{1}{k-2}, \frac{2 S^{(-k+2)} r}{(k-2)\sigma^2}\right)\left(\frac{2 S^{(-k+2)} r}{(k-2)\sigma^2}\right)^{\left(\frac{k}{k-2}\right)} e^{\left(-\frac{2 S^{(-k+2)} r}{(k-2)\sigma^2}\right)} dS}{2^{\left(\frac{1}{k-2}\right)} r^{\left(\frac{1}{k-2}\right)} \sigma^{\left(-\frac{2}{k-2}\right)} (k-2)^{\left(-\frac{k-1}{k-2}\right)} \left(\frac{m k - 2m + 1}{k-2}\right)! e^{(r(mk-2m+1)T)}} = c_m$$

(26)

Replacing (26) in (17), we obtain

$$V(t,S) = \sum_{n=0}^\infty \left( \frac{2^{\left(\frac{k}{k-2}\right)} n! \, \text{LaguerreL}\left(n, \frac{1}{k-2}, \frac{2 S^{(-k+2)} r}{(k-2)\sigma^2}\right) e^{(-r(nk-2n+1)(-t+T))}}{2^{\left(\frac{1}{k-2}\right)} r^{\left(\frac{1}{k-2}\right)} \sigma^{\left(-\frac{2}{k-2}\right)} (k-2)^{\left(-\frac{k-1}{k-2}\right)} \left(\frac{nk-2n+1}{k-2}\right)!} \right.$$

$$\left. \int_0^\infty \text{LaguerreL}\left(n, \frac{1}{k-2}, \frac{2 \Sigma^{(-k+2)} r}{(k-2)\sigma^2}\right) \left(\frac{\Sigma^{(-k+2)} r}{(k-2)\sigma^2}\right)^{\left(\frac{k}{k-2}\right)} d\Sigma \, e^{\left(-\frac{2 \Sigma^{(-k+2)} r}{(k-2)\sigma^2}\right)} \frac{A \Sigma^{(p+1)} e^{(-\alpha \Sigma)} \alpha^{(p+1)}}{\Gamma(p+1)} \right)$$

(27)

which is the solution of (5) for the Aunt Michaela option (18).

## References

1. Liviu-Adrian Cotfas, Camelia Delcea, Nicolae Cotfas, Exact solution of a generalized version of the Black-Scholes equation, arXiv:1411.2628.

2. Natanael Karjanto, Binur Yermukanova, Laila Zhexembay , Black-Scholes equation , arXiv:1504.03074

3. Y. Zheng, On Generalized Stochastic Differential Equation and Black-Scholes Dynamic Process, Proceedings of the World Congress on Engineering 2010 Vol I WCE 2010, June 30 - July 2, 2010, London, U.K.;
http://www.iaeng.org/publication/WCE2010/WCE2010_pp364-367.pdf

**3.A.** Ospina Juan, New Analytical Solutions of a Modified Black-Scholes Equation with the European Put Option, https://arxiv.org/ftp/arxiv/papers/1508/1508.03841.pdf

4. Maple, www.maplesoft.com

5. Kummer functions,

https://en.wikipedia.org/wiki/Confluent_hypergeometric_function

6. Associated Laguerre Polynomials,

https://en.wikipedia.org/wiki/Laguerre_polynomials

7. Orhtonomalization for associated Laguerre polynomials,

http://mathworld.wolfram.com/AssociatedLaguerrePolynomial.html